\documentclass[twocolumn,aps,pra,longbibliography]{revtex4-1}

\usepackage{natbib}

\usepackage{times}
\usepackage{graphicx}
\usepackage{amsfonts}
\usepackage{amsmath, amsthm, amssymb}
\usepackage{empheq}
\usepackage{dsfont}
\usepackage{color}
\usepackage{mathdots}
\usepackage{fancyvrb}
\usepackage{standalone}
\usepackage{microtype}
\usepackage{tikz}
\usetikzlibrary{decorations.markings}
\usetikzlibrary{shapes,positioning,arrows}
\usepackage{bm}
\usepackage{siunitx}
\usepackage[colorlinks=true, allcolors=black, citecolor=blue, linkcolor=blue, urlcolor=blue]{hyperref}
\usepackage[capitalize]{cleveref}
\allowdisplaybreaks

\definecolor{mygray}{RGB}{200, 200, 200}

\newcommand{\s}{\sigma}

\newcommand{\barm}{\bar{m}}

\renewcommand{\L}{\mathcal{L}}
\newcommand{\D}{\mathcal{D}}
\newcommand{\G}{\mathcal{G}}

\DeclareMathOperator{\Tr}{Tr}
\renewcommand{\det}{\mathrm{det}~}

\newcommand{\del}{\partial}

\newcommand{\vk}{{\bm{\mathrm{k}}}}
\newcommand{\vK}{{\bm{\mathrm{K}}}}
\newcommand{\vq}{{\bm{\mathrm{q}}}}
\newcommand{\vQ}{{\bm{\mathrm{Q}}}}

\newcommand{\vP}{{\bm{\mathrm{P}}}}
\newcommand{\vm}{{\bm{\mathrm{m}}}}

\renewcommand{\vr}{{\bm{\mathrm{r}}}}
\newcommand{\vf}{v_\rm{F}}
\newcommand{\kf}{k_\rm{F}}

\newcommand{\va}{{\bm{a}}}
\newcommand{\vn}{{\bm{n}}}
\newcommand{\bnabla}{\bm{\nabla}}
\newcommand{\M}{\mathcal{M}}

\newcommand{\on}{\omega_n}
\newcommand{\On}{{\Omega_n}}

\renewcommand{\v}[1]{{\bm{#1}}}
\newcommand{\vs}{\bm{\s}}
\newcommand{\eps}{\epsilon}

\newcommand{\su}{\uparrow}
\newcommand{\sd}{\downarrow}

\renewcommand{\rm}[1]{\mathrm{#1}}

\makeatother

\usepackage{soul}

\allowdisplaybreaks[1]

\begin{document}

\title{Possible odd-frequency Amperean magnon-mediated superconductivity in topological insulator -- ferromagnetic insulator bilayer}
\author{Henning G. Hugdal}
\author{Asle Sudb{\o}}
\affiliation{Center for Quantum Spintronics, Department of Physics, NTNU, Norwegian University of Science and Technology, NO-7491 Trondheim, Norway}

\begin{abstract}
We study the magnon-mediated pairing between fermions on the surface of a topological insulator (TI) coupled to a ferromagnetic insulator with a tilted mean field magnetization. Tilting the magnetization towards the interfacial plane reduces the magnetic band gap and leads to a shift in the effective TI dispersions. We derive and solve the self-consistency equation for the superconducting gap in two different situations, where we neglect or include the frequency dependence of the magnon propagator. Neglecting the frequency dependence results in $p$-wave Amperean solutions. We also find that tilting the magnetization into the interface plane favors Cooper pairs with center of mass momenta parallel to the magnetization vector, increasing $T_c$ compared to the out-of-plane case. Including the frequency dependence of the magnon propagator, and solving for a low number of Matsubara frequencies, we find that the eigenvectors of the Amperean solutions at the critical temperature are dominantly odd in frequency and even in momentum, thus opening the possibility for odd-frequency Amperean pairing.
\end{abstract}

\maketitle

\section{Introduction}
Spin fluctuations is one of the proposed mechanisms for superconductivity in unconventional superconductors \cite{Scalapino2012,Stewart2017}, for which the phase diagrams often have both antiferromagnetic and superconducting regions \cite{Moriya1990,Monthoux1991,Monthoux1992,Monthoux1992b,Moriya2000,Moriya2003,Moriya2006}, or where ferromagnetism and superconductivity appear simultaneously \cite{Kirkpatrick2001,Suhl2001,Kirkpatrick2003,Karchev2003,Kar2014,Funaki2014}. Recently, there have been studies focusing on the possibility of magnon-mediated superconductivity in heterostructures consisting of magnetic insulators and a normal metal or topological insulator (TI) \cite{Kargarian2016,Gong2017,Rohling2018,Hugdal2018b,Erlandsen2019,Fjaerbu2019,Erlandsen2020}, where the electrons couple to magnetic fluctuations at the interface. In TIs the superconductivity can be between fermions with parallel momenta, so-called Amperean pairing \cite{Lee2007}. It has also been shown that a coupling to magnons can lead to indirect exciton condensation \cite{Johansen2019}.

Coupling the magnetic insulator to the TI surface states \cite{Hasan2010,Qi2011} has a few interesting consequences compared to coupling to the electrons in a normal metal. First of all, the metallic states are restricted to the surface, locating them close to the spin fluctuations, ensuring a strong coupling. Moreover, due to the spin-momentum locking in the TI, the response to the magnetization is very different compared to a normal metal. While an exchange field leads to a band splitting and thus pair-breaking effects for any spin-$0$ Cooper pairs in a normal metal, the exchange field in a TI leads only to a gap and/or shift in the surface state dispersions, but no band splitting. Hence, the Fermi level only crosses one band, and the Cooper pairs must necessarily be pseudo-spin triplets.

In this work we study a TI exchange coupled to a ferromagnetic insulator (FMI) with a mean field magnetization that can be tilted towards the plane of the interface between the TI and FMI. We derive the gap equation for the static gap, and study the possibility of both Bardeen-Cooper-Schrieffer (BCS) \cite{Bardeen1957} type superconductivity and Amperean superconductivity, focusing on the changes due to the in-plane component of the magnetization. We also derive the gap equation including the frequency dependence of the magnon propagator, and solve these equations including only a few Matsubara frequencies. Our results show that the eigenvectors are mostly odd in frequency \cite{Berezinskii1974a,*Berezinskii1974,Balatsky1992,Coleman1993,Coleman1993a,Coleman1994,Abrahams1995,Bergeret2005,Tanaka2012,Linder2019}, thus showing the possibility for magnon-mediated \emph{odd-frequency Amperean superconductivity.}

The remainder of the article is organized as follows: The model is presented in Sec.~\ref{sec:model}, as is the derivation of the effective magnon-mediated action. The general gap equations are derived in Sec.~\ref{sec:gapequations}, and specifically studied for the static and frequency dependent cases in \cref{sec:static,sec:fullfreq}. Finally, the main results are summarized in Sec.~\ref{sec:conclusion}. Further details regarding the derivations and material parameters are presented in the Appendix.

\section{Model}\label{sec:model}
\begin{figure}
	\includegraphics[width=0.7\columnwidth]{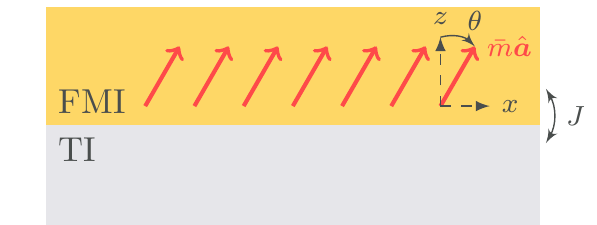}
	\caption{\label{fig:system} Sketch of the system consisting of TI coupled to a FMI with a mean field magnetization tilted in the $xz$ plane by angle $\theta$ with respect to the $z$ axis. $J$ is the strength of the exchange coupling.}
\end{figure}

A sketch of the system is shown in \cref{fig:system}. We model the FMI using the Lagrangian
\begin{align}
    \L_m = -\v{b}(\vm)\cdot \del_t \vm- \frac{\kappa}{2}(\bm{\nabla} \vm)^2 + \lambda (\vm\cdot\hat{\va})^2,\label{eq:Lm_t}
\end{align}
where $\hat{\va}$ is the direction of the mean field magnetization, parametrized by $\hat{\va} = \sin\theta \hat{x} + \cos\theta\hat{z}$, and $\lambda > 0$. A general mean field magnetization including a $y$ component can be shown to be equivalent to considering only a $xz$ plane magnetization by rotating the spin-quantization axis and the coordinate system. $\v{b}(\vm)$ is the Berry connection, satisfying $\bnabla_\vm \times \v{b}(\vm) = \vm/\barm^2$ \cite{Auerbach1994}, where $\bnabla_\vm = (\partial_{m_x}, \partial_{m_y}, \partial_{m_z})$. Here $\barm$ is the length of the mean field magnetization along $\hat{\va}$. We have set $\hbar=1$ throughout the paper. The Lagrangian of the TI surface states reads
\begin{align}
    \L_\mathrm{TI} = \Psi^\dagger\left[i\del_t - i \vf (\s_y\del_x - \s_x \del_y) + \mu\right]\Psi, \label{eq:L_TI_realt}
\end{align}
where $\Psi = (\psi_\su,~\psi_\sd)^T$ is a vector of spin up and spin down electrons in the TI, $\vf$ is the Fermi velocity, and $\mu$ is the chemical potential. The TI and FMI are coupled via the exchange coupling term
\begin{align}
    \L_c = J\Psi^\dagger \vm\cdot\vs \Psi,
\end{align}
where $J$ is the coupling strength.

We fix the length of the magnetization vector to $\barm$, and thus write
\begin{align}
    \vm = \sqrt{1 - \vn^2}\barm \hat{\va} + \barm\vn,
\end{align}
where $\vn$ is the fluctuation vector perpendicular to $\hat{\va}$, 
\begin{align}
    \vn = n(\cos\theta \hat{x} - \sin\theta \hat{z}) + n_y \hat{y}.
\end{align}
We assume that $n,n_y \ll 1$.

We next calculate the Berry connection by generalizing the leading order expression $\v{b} = (\hat{z}\times \vn)/2$ \cite{Kargarian2016} to a mean field direction along $\hat{\va}$:
\begin{align}
    \v{b} = \frac{\hat{\va}\times\vn}{2} = -\frac{n_y \cos\theta \hat{x} - n \hat{y} - n_y\sin\theta\hat{z}}{2}.
\end{align}
Hence, to lowest order, we have
\begin{align}
    \bnabla_\vm\times \v{b} = \frac{\hat{\va}}{\barm}.
\end{align}

Switching to imaginary time $\tau=it$ and Fourier transforming \footnote{We use the convention
\begin{align*}
	f(\tau,\vr) = \frac{1}{\beta V}\sum_{\on,\vk} f(\on,\vk) e^{i\vk\cdot\vr - i\on\tau}
\end{align*} for the Fourier transform.}, we get the three contributions to the action
\begin{align}
    S_m ={}& \frac{1}{\beta V} \sum_q \bigg\{\left[\frac{\kappa\barm^2}{2}\vq^2+\lambda\barm^2\right]\vn(-q)\cdot \vn(q) \nonumber\\*
    &- \frac{\On\barm}{2}\left[n_y(-q)n(q) - n(-q)n_y(q)\right]\bigg\},\\
    S_\mathrm{TI} ={}& \frac{1}{\beta V} \sum_k \Psi^\dagger(k)[-i\on - \vf(k_x\s_y - k_y \s_x) - \mu]\Psi(k),\\
    S_c ={}& S_c^{\barm} + S_c^\vn \nonumber\\*
    ={}& -\frac{J\barm}{\beta V} \sum_{k} \Psi^\dagger(k) \hat{\va}\cdot\vs \Psi(k) \nonumber\\
    &-\frac{J\barm}{(\beta V)^2} \sum_{q,k} \Psi^\dagger(k+q) \vn(q)\cdot \vs \Psi(k).
\end{align}
Here we have used the notation $q=(\On,\vq)$ and $k=(\on,\vk)$ for bosonic and fermionic Matsubara frequencies and momenta respectively. We have also kept only leading order terms in the fluctuations in the coupling term. Using a more general model as a starting point, such as the one in Ref.~\cite{Rex2017}, $\lambda$ could in principle be renormalized to take negative values, meaning that an antiferromagnetic alignment between the magnetic fluctuations $\vn$ could be favored.

\subsection{Integrating out the magnons}
To obtain the effective, magnon-mediated interaction between Dirac electrons, we need to integrate out the magnons. This can be done by rewriting the full magnon action $S_\vn = S_m + S_c^\vn$ by introducing the vectors $N(q) = (n(q), ~n_y(q))^T$ and
\begin{align}
    j(q) = \frac{J\barm}{\beta V}\sum_k \begin{pmatrix}
        \Psi^\dagger(k+q)(\cos\theta \s_x - \sin\theta\s_z)\Psi(k)\\
        \Psi^\dagger(k+q)\s_y\Psi(k)
    \end{pmatrix},
\end{align}
resulting in
\begin{align}
    S_\vn ={}& \frac{1}{\beta V}\sum_q \bigg\{\nonumber\\*
    &N(-q)^T\left[\frac{\kappa\barm^2}{2}\vq^2+\barm^2\lambda +\frac{i\On\barm\s_y}{2}\right]N(q) \nonumber\\
    &- \frac{N^T(-q)j(-q) + j^T(q)N(q)}{2}\bigg\}.
\end{align}
Performing the functional integral, we get an additional term in the TI action,
\begin{align}
    \delta S_\mathrm{TI} = -\frac{1}{4\beta V\barm}\sum_q j^T(q) \frac{\frac{\kappa\barm}{2}\vq^2 + \barm\lambda -4 \frac{i\On}{2}\s_y}{\left(\frac{\On}{2}\right)^2 + \left(\frac{\kappa\barm}{2}\vq^2 + \barm\lambda\right)^2}j(-q).
\end{align}
In the low frequency limit, the last term in the numerator is less singular than the other two terms, and we therefore neglect it in the following \cite{Kargarian2016}. We therefore get
\begin{align}
    \delta S_\mathrm{TI} = -\frac{1}{4\beta V\barm}\sum_q\frac{\omega_\vq}{\left(\frac{\On}{2}\right)^2 + \omega_\vq^2}j^T(q)j(-q)\label{eq:deltaS_TI},
\end{align}
where we have defined the magnon dispersion
\begin{align}
    \omega_\vq = \frac{\kappa\barm}{2}\vq^2 + \barm\lambda. \label{eq:omega}
\end{align}

\subsection{Diagonalization of mean field TI action}
\begin{figure}
    \centering
    \includegraphics[width=\columnwidth]{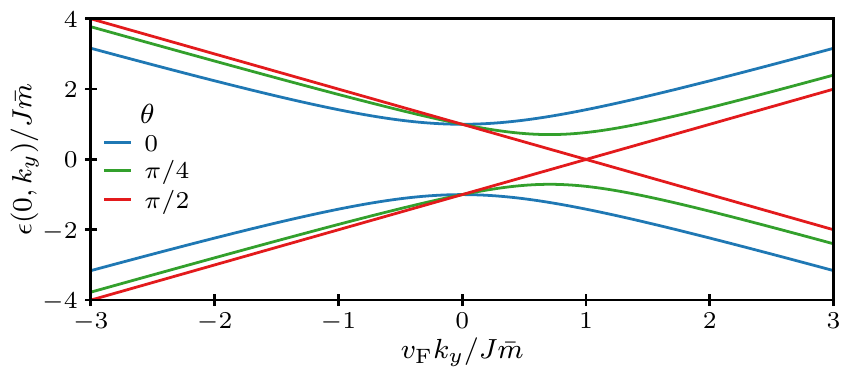}
    \caption{\label{fig:dispersion} Plot of eigenenergies in \cref{eq:epsilon} as a function of $k_y$ with $k_x=0$ and $\mu=0$ for different values of $\theta$. Increasing $\theta$ towards $\pi/2$ reduces the mass gap and shifts the center of the dispersion away from $k_y=0$. At $\pi/2$ we have a Dirac point located at $k_y = J\barm/\vf$.}
\end{figure}

We next diagonalize the mean field TI action,
\begin{align}
    S_\mathrm{TI}^\mathrm{mf} = -\frac{1}{\beta V}\sum_k \Psi^\dagger(k)G^{-1}(k)\Psi,
\end{align}
where we have defined the inverse Green's function
\begin{align}
    G^{-1}(k) ={}& i\on + \mu + M\s_z + \vf k_x \s_y - \vf(k_y - K_y)\s_x,
\end{align}
where $M = J\barm \cos\theta$ and $K_y = J\barm\sin\theta/\vf$.
Diagonalizing the Green's function results in
\begin{align}
    G_d^{-1} = P_\vk G^{-1} P_\vk^\dagger = \operatorname{diag}(\lambda_+,\lambda_-),
\end{align}
where the diagonal entries are
\begin{align}
    \lambda_\pm = i\on + \mu \mp \sqrt{\vf^2k_x^2+\vf^2(k_y-K_y)^2 + M^2},
\end{align}
and the Green's function is diagonalized by the matrix
\begin{align}
    P_\vk = \frac{1}{\sqrt{n_\vk}}\begin{pmatrix}
        s_\vk^* & r_\vk\\
        -r_\vk & s_\vk
    \end{pmatrix},
\end{align}
where
\begin{subequations}
\begin{align}
    s_\vk ={}& \vf(k_y - K_y) + i \vf k_x,\\
    r_\vk ={}& M + \sqrt{|s_\vk|^2+M^2},\\
    n_\vk ={}& r_\vk^2 + |s_\vk|^2.
\end{align}
\end{subequations}
The eigenvectors $\Psi_\pm(k)$ in the helicity basis are given by a transformation from the spin basis $\Psi(k)$, defined below \cref{eq:L_TI_realt}, as follows,
\begin{align}
    \Psi_\pm(k) \equiv \begin{pmatrix} \psi_+\\ \psi_-\end{pmatrix}= P_\vk\Psi(k), \label{eq:dirac_fermions}
\end{align}
where the helicity index is denoted by $+$ or $-$. The eigenenergies are given by the zeros of the diagonal entries,
\begin{align}
    \eps_\pm(\vk) = \pm \sqrt{\vf^2k_x^2+\vf^2(k_y-K_y)^2 + M^2} - \mu.\label{eq:epsilon}
\end{align}
Hence,  $M$ leads to a gap in the dispersion, while $K_y$ shifts the dispersion along the $k_y$ axis. This is illustrated in \cref{fig:dispersion}.

\subsection{Magnon-mediated interaction}
We now rewrite the effective action in \cref{eq:deltaS_TI} in terms of the Dirac fermions defined by \cref{eq:dirac_fermions}, assuming that $\mu > M$ and thus restricting the problem to only considering the $\psi_+$ fermions. This results in (see \cref{app:interaction} for details)
\begin{align}
    \delta S_\mathrm{TI} ={}& -\frac{J^2\barm}{4(\beta V)^3} \sum_{q,k,k'} D(q)\Lambda_{\vk'\vk}(\vq) \nonumber\\*
    &\times \psi^\dagger(k'+q) \psi^\dagger(k-q) \psi(k)\psi(k'),
\end{align}
where we for notational simplicity have dropped the subscript $+$ on the fields $\psi_+$, and defined the magnon propagator
\begin{align}
    D(q) = \frac{\omega_\vq}{(\On/2)^2 + \omega_\vq^2}, \label{eq:D}
\end{align}
and the scattering form factor $\Lambda_{\vk'\vk}(\vq) = \Lambda^0_{\vk'\vk}(\vq) + \Lambda^{x}_{\vk'\vk}(\vq) + \Lambda^{xz}_{\vk'\vk}(\vq)$, with
\begin{align}
    \Lambda^0_{\vk'\vk}(\vq) ={}& \frac{\cos^2\theta + 1}{\sqrt{n_\vk n_{\vk-\vq} n_{\vk'} n_{\vk'+\vq}}} \nonumber\\*
    &\times\Big[s_{\vk'}s_{\vk-\vq}^*r_{\vk'+\vq}r_\vk + s_\vk s_{\vk'+\vq}^*r_{\vk'}r_{\vk-\vq}\Big]\\
    \Lambda^x_{\vk'\vk}(\vq) ={}& \frac{\sin^2\theta}{\sqrt{n_\vk n_{\vk-\vq} n_{\vk'} n_{\vk'+\vq}}} \nonumber\\*
    &\times \Big[s_{\vk'}s_\vk s_{\vk'+\vq}^* s_{\vk-\vq}^*  - s_{\vk'}s_{\vk'+\vq}^*r_{\vk}r_{\vk-\vq} \nonumber\\*
    &- s_\vk s_{\vk-\vq}^* r_{\vk'} r_{\vk'+\vq} + r_\vk r_{\vk-\vq} r_{\vk'}r_{\vk'+\vq} \nonumber\\*
    &- s_{\vk'}s_\vk r_{\vk'+\vq}r_{\vk-\vq} - s_{\vk'+\vq}^* s_{\vk-\vq}^* r_{\vk'}r_\vk\Big]\\
    \Lambda^{xz}_{\vk'\vk}(\vq) ={}& -\frac{\cos\theta\sin\theta}{\sqrt{n_\vk n_{\vk-\vq} n_{\vk'} n_{\vk'+\vq}}} \nonumber\\*
    &\times \Big[s_{\vk'}s_\vk s_{\vk-\vq}^* r_{\vk'+\vq} + s_\vk s_{\vk'} s_{\vk'+\vq}^* r_{\vk-\vq} \nonumber\\*
    &+ s_{\vk} s_{\vk-\vq}^* s_{\vk'+\vq}^* r_{\vk'}  + s_{\vk'}s_{\vk'+\vq}^* s_{\vk-\vq}^* r_\vk\nonumber\\*
    &- s_{\vk'}r_{\vk'+\vq} r_{\vk} r_{\vk-\vq}  - s_\vk r_{\vk-\vq} r_{\vk'} r_{\vk'+\vq} \nonumber\\*
    &- s_{\vk'+\vq}^*r_{\vk'}r_{\vk}r_{\vk-\vq} - s_{\vk-\vq}^*r_{\vk}r_{\vk'} r_{\vk'+\vq}\Big].
\end{align}
The first expression is the same expression as was analyzed in Refs.~\cite{Kargarian2016,Hugdal2018b}, except it now has a $\theta$-dependence and an overall multiplicative factor of 2 when $\theta = 0$. This term is, however, always non-zero. The other two expressions were not present in Refs.~\cite{Kargarian2016,Hugdal2018b}, as they both require an $x$-component in the mean field magnetization. The last expression also requires a finite $z$-component. Hence we may have differences in the pairing depending on the angle of the mean field direction, which will be analyzed after calculating the gap equations for the system.

\section{Gap equations}\label{sec:gapequations}
Including the symmetrized magnon mediated interaction in the action, we get the following effective action for the $+$ fermions,
\begin{align}
    S_+ &= -\frac{1}{\beta V}\sum_k \psi^\dagger(k)\lambda_+(k) \psi(k) + \frac{1}{(\beta V)^3}\sum_{k,k',q} V_{k'k}(q) \nonumber\\*
    &\times\psi^\dagger\left(k'+\frac{q}{2}\right) \psi^\dagger\left(-k'+\frac{q}{2}\right) \psi\left(-k+\frac{q}{2}\right) \psi\left(k+\frac{q}{2}\right),
\end{align}
with the symmetrized interaction
\begin{align}
    V_{k'k}(q) ={}& -\frac{J^2\barm}{8}\bigg[D(k'-k) \Lambda_\vq(\vk',\vk) \nonumber\\*
    &- D(k'+k) \Lambda_\vq(\vk',-\vk)\bigg].
\end{align}
For notational simplicity we have defined
\begin{align}
    \Lambda_{\vq}(\vk',\vk) \equiv \Lambda_{\vk+\frac{\vq}{2},-\vk+\frac{\vq}{2}}(\vk'-\vk).
\end{align}
We have also relabeled the momenta to allow for a finite center of mass momentum $\vq$ for the Cooper pairs, which is necessary for Amperean pairing. Moreover, since the minimum of the dispersion is shifted away from $\vk=0$ for non-zero $\theta$ there is also the possibility of BCS Cooper pairs with finite center of mass momentum, i.e. a Fulde-Ferrell-Larkin-Ovchinnikov (FFLO) state \cite{Fulde1964,Larkin1965}. As such, the system has some similarities to two-dimensional normal metal systems with Rashba spin-orbit coupling coupled to a Zeemann field with an in-plane component, leading to a shift in the dispersion and thus the possibility of a FFLO state \cite{Barzykin2002,Dimitrova2007,Agterberg2007,Loder2013,Lake2016,Hugdal2018}.

We now perform a Hubbard-Stratonovich decoupling \cite{Altland2010} by introducing bosonic fields $\varphi_q$ and $\phi^\dagger_q$ (see Appendix \ref{app:HSdecoupling} for details), resulting in the functional integral
\begin{align}
    Z = {}& \int \D\psi^\dagger\D\psi ~ e^{-S'}\int\D\varphi_q^\dagger\D\varphi_q ~ e^{-S_\phi^0},\label{eq:Z}
\end{align}
where we have the fermionic action containing the coupling to the bosonic fields
\begin{align}
    S' = {}&-\frac{1}{\beta V}\sum_k\bigg\{\psi^\dagger(k)\lambda_+(k)\psi(k) \nonumber\\*
    &+\sum_q\Big[\varphi_q^\dagger(k)\psi\left(-k+\frac{q}{2}\right) \psi\left(k+\frac{q}{2}\right) \nonumber\\*
    &+ \psi^\dagger\left(k+\frac{q}{2}\right) \psi^\dagger\left(-k+\frac{q}{2}\right)\varphi_q(k)\Big]\bigg\},
\end{align}
and the additional bosonic action
\begin{align}
    S_\phi^0 ={}& - \beta V\sum_{q,k'k}\varphi_q^\dagger(k')[V_{k'k}(q)]^{-1}\varphi_q(k).
\end{align}
Before proceeding any further, we will assume that the mean field bosonic field is of the form
\begin{align}
    \varphi_q(k) = \frac{1}{2}\delta_{\vq,\vQ}\delta_{\On,0}\Delta_\vQ(k).
\end{align}
This effectively restricts the analysis to only consider Cooper pairs with one common center of mass momentum. In general, these will couple to Cooper pairs with other center of mass momenta. However, since any interaction between them does not conserve momentum, the couplings are likely to be small, and we therefore focus on only one $\vQ$ in the following.

In order to integrate out the fermions, we rewrite the action using the vector
\begin{align}
    \Psi_Q(k) = \begin{pmatrix}
        \psi(k)\\
        \psi^\dagger(-k+Q)
    \end{pmatrix},
\end{align}
where $Q=(0,\vQ)$, leading to
\begin{align}
    S' = -\frac{1}{2\beta V}\sum_k \Psi_Q^\dagger\left(k+\frac Q2 \right)\G^{-1}_Q(k)\Psi_Q\left(k+\frac Q2\right)
\end{align}
where we have defined the inverse Green's function matrix
\begin{align}
\G^{-1}_Q(k) = \begin{pmatrix} 
        \lambda_+\left(k+\frac Q2 \right) & \Delta_\vQ(k)\\
        \Delta_\vQ^\dagger(k) & -\lambda_+\left(-k+\frac Q2\right)
\end{pmatrix}.
\end{align}
Integrating out the fermions, we finally get the effective action for the bosonic fields
\begin{align}
    S_\phi = {}&- \frac{\beta V}{4}\sum_{k'k}\Delta_\vQ^\dagger(k')[V_{k'k}(Q)]^{-1}\Delta_\vQ(k) \nonumber\\*
    &- \frac{1}{2}\Tr\ln(-\G_Q^{-1}).
\end{align}

The gap equation follows from using the saddle point approximation \cite{Altland2010},
\begin{align}
    \frac{\delta S_\phi}{\delta \Delta_\vQ(p)} = 0,
\end{align}
resulting in
\begin{align}
    \frac{\beta V}{4} \sum_{k'}{}&\Delta_\vQ^\dagger(k')[V_{k'p}(Q)]^{-1} = \frac{\Delta_\vQ^\dagger(p)}{2\det \G_Q^{-1}(p)}
\end{align}
where $\det \G_Q^{-1}(k) = -\lambda_+(k+Q/2)\lambda_+(-k+Q/2) - |\Delta_\vQ(k)|^2$.
Multiplying both sides with $V_{pk}(Q)/\beta V$ and summing over $p$, we get
\begin{widetext}
\begin{align}
     \Delta_\vQ^\dagger(k) = \frac{2}{\beta V} \sum_{\on',\vk'} \frac{\Delta^\dagger_\vQ(k')V_{k'k}(Q)}{[i\on' - \eps_\vQ^o(\vk') - E_\vQ(k')][i\on' - \eps_\vQ^o(\vk') + E_\vQ(k')]}. \label{eq:basic_gap_eq}
\end{align}
\end{widetext}
where we have defined
\begin{align}
    \eps_\vQ^o(\vk') ={}& \frac{\eps_+\left({\vk'+\frac{\vQ}{2}}\right) - \eps_+\left({-\vk'+\frac{\vQ}{2}}\right)}{2},\\
    \eps_\vQ^e(\vk') ={}& \frac{\eps_+\left({\vk'+\frac{\vQ}{2}}\right) + \eps_+\left({-\vk'+\frac{\vQ}{2}}\right)}{2},\\
    E_\vQ(k') ={}& \sqrt{[\eps_\vQ^e(\vk')]^2 + |\Delta_\vQ(k')|^2}.
\end{align}
{It is important to point out that since we have included only the $\psi_+$ states in the analysis, the gap equation is for pseudo-spin triplets, the "spin" in this case being the helicity index $+$ or $-$. The physical spin is not a good quantum number because of the spin-orbit coupling in the system. Therefore, following the symmetry analysis in e.g. Ref.~\cite{Linder2019}, the gap function has to be odd in $\on$ and even in $\vk$, or even in $\on$ and odd in $\vk$.}

We will now treat the gap equation in two different ways: (1) We neglect the frequency dependence of the magnon propagator in \cref{eq:D} \cite{Kirkpatrick2003,Karchev2003,Hugdal2018b,Johansen2019} and study the static limit, and (2) we use an approach similar to the Eliashberg equations \cite{Eliashberg1960a,*Eliashberg1960b,Eliashberg1960c,*Eliashberg1961,Mahan2000}, solving the gap equations directly including only a low number of Matsubara frequencies.

\section{Frequency independent solution}\label{sec:static}
In the static limit, we set the frequency to zero in the magnon propagator,
\begin{align}
    D(q) \to D(\vq) = \frac{1}{\omega_\vq},
\end{align}
such that the interaction now only depends on the momenta, $V_{k'k}(q) \to V_{\vk'\vk}(\vq)$. Hence, there is no longer a free frequency in the gap equation, and we can perform the remaining Matsubara sum, resulting in
\begin{align}
    \Delta_\vQ^\dagger(\vk) ={}& -\frac{2}{V} \sum_{\vk'} V_{\vk'\vk}(\vQ)\Delta_\vQ(\vk')\chi_\vQ(\vk'),
\end{align}
with
\begin{align}
    \chi_\vQ(\vk) ={}& \frac{1}{4E_\vQ(\vk')}\bigg[\tanh\frac{\beta\left(\eps_\vQ^o(\vk)+E_\vQ(\vk)\right)}{2} \nonumber\\*
    &- \tanh\frac{\beta\left(\eps_\vQ^o(\vk)-E_\vQ(\vk)\right)}{2}\bigg],
    \label{eq:chi}
\end{align}
where $\beta = 1/k_\rm{B} T$. Since the surface states are pseudo-spin triples, and we have neglected the frequency dependence, $\Delta_\vQ(\vk)$ must now be an odd function of $\vk$ \cite{Linder2019}, which is evident also from the interaction. For simplicity, we now define $\vK = (0, K_y)$, and let $\vQ = 2\vK + 2\vP$, such that the center of the Fermi surface is at the origin when $\vP = 0$ independent of the angle $\theta$.

\subsection{BCS pairing}
We first study the case $\vP=0$, which resembles the regular BCS pairing case with circular Fermi surface. Now $\eps_{2\vK}^o(\vk) = 0$ for all $\vk$, and the temperature dependent factor in the gap equation simplifies to
\begin{align}
    \chi_{2\vK}(\vk) = \frac{1}{2E_{2\vK}(\vk)}\tanh\frac{\beta E_{2\vK}(\vk)}{2},
\end{align}
which is peaked at the minima of $E_{2\vK}$, at Fermi momenta $\vf \kf  = \sqrt{\mu^2-M^2}$. Instead of solving the gap equation directly, we write the linearized gap equation \cite{Sigrist2005}
\begin{align}
    \Delta_{2\vK}^\dagger(\vk) ={}& -\left<2V_{\vk'\vk}(2\vK)\Delta_{2\vK}(\vk')\right>_{\vk',\rm{FS}} \int \frac{dk'}{2\pi}k'\chi_{2\vK}(k'),
\end{align}
which can be written as an eigenvalue problem
\begin{align}
    \eta \Delta_{2\vK}(\vk') = -\left<2V_{\vk'\vk}(2\vK)\Delta_{2\vK}(\vk')\right>_{\vk',\rm{FS}}, \label{eq:BCS_eigenvalue_eq}
\end{align}
where FS denotes an average over the Fermi surface. The critical temperature is then proportional to $e^{-c/\eta}$, where $\eta$ is the highest positive eigenvalue \cite{Sigrist2005}, and $c$ is some constant.

\begin{figure}
\includegraphics{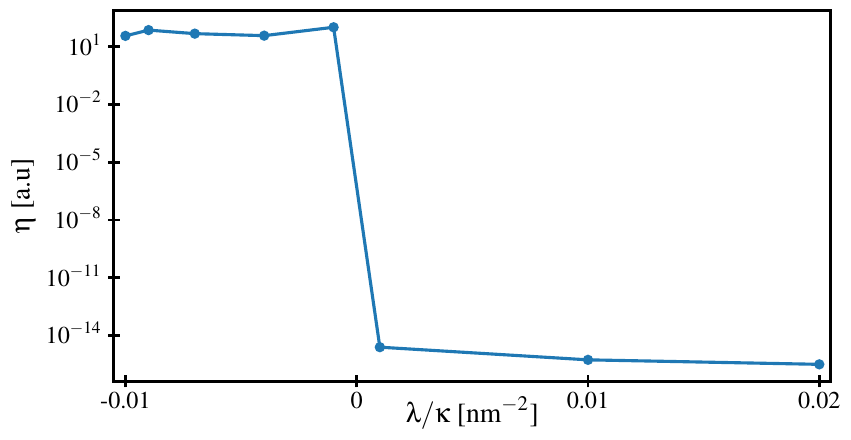}
\caption{\label{fig:BCS_eigenvalues} Plot of solutions to the eigenvalue problem \cref{eq:BCS_eigenvalue_eq} as a function of $\lambda$ for $\theta=0$ and $J = \SI{0.01}{eV nm^2}$. We see that the eigenvalues are very small for $\lambda > 0$, indicative of the gap equation not having solutions. 
For $\lambda < 0$ however, we get finite eigenvalues, meaning that the gap equation has solutions for negative $\lambda$.
}    
\end{figure}

Assuming $M \ll \mu$, we get for the scattering form factor
\begin{align}
    \Lambda_{2\vK}(\vk',\vk) \approx - e^{-i\phi_\vk + i\phi_{\vk'}}\left[1 - \sin^2\theta\sin\phi_\vk\sin\phi_{\vk'}\right].\label{eq:Lambda_BCS_approx}
\end{align}
The expression in the square bracket never changes sign, but introduces anisotropy in $\vk$-space. Hence, for $\phi_\vk = \phi_{\vk'}$, the interaction $V_{\vk'\vk}(2\vK)$ is always positive as long as the easy axis parameter $\lambda>0$ (see \cref{eq:Lm_t}), giving the wrong overall sign in order for a non-trivial solution of the gap equation to be possible. To verify this, we solve \cref{eq:BCS_eigenvalue_eq} numerically as a function of $\lambda$ using the parameter values in \cref{tab:parameters}, resulting in the eigenvalues shown in \cref{fig:BCS_eigenvalues} for tilt angle $\theta=0$. The figure shows that $\eta$ is very small for positive values of $\lambda$. We also calculated the eigenvectors, which were randomly fluctuating for positive $\lambda$. For other tilt angles, the results are qualitatively the same. Hence, we conclude that BCS pairing is not possible in the static limit for $\lambda > 0$. For $\lambda <0$ we find finite eigenvalues $\eta$ and smooth eigenvectors, meaning that the system has a superconducting instability in this case, the reason being that the magnon propagator, and thus the interaction potential, now can change sign. This is consistent with the results in Ref.~\cite{Hugdal2018b}. In systems where $\lambda <0$ and $\theta \neq 0$ is possible, this would lead to FFLO Cooper pairs with momentum $2\vK$. However, for the present system, we have assumed that $\lambda>0$, thus we do not find a solution to the gap equation in the BCS like case.

\subsection{Amperean pairing}\label{subsec:amperean}
\begin{figure}
    \includegraphics{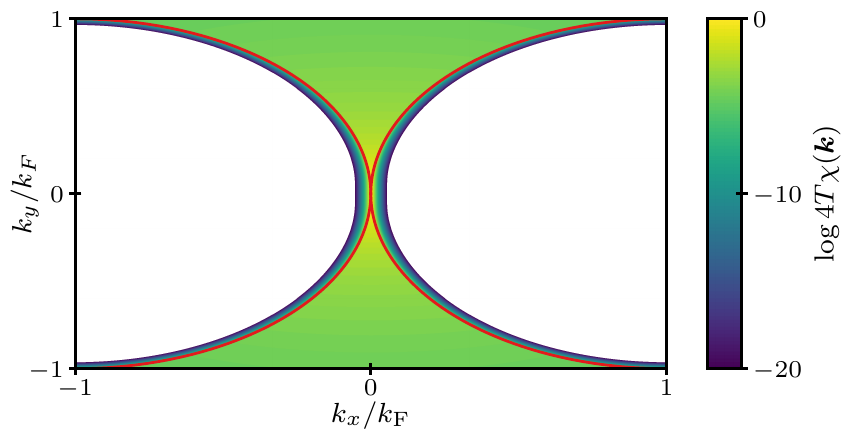}
    \caption{\label{fig:chi} Plot of the logarithm of $4T\chi_{\vQ}(\vk)$ with $\vQ = 2\vK + 2\vP$ for $\vP=(\kf,0)$, $k_\rm{B} T = 5\times10^{-4}\si{eV}$ and $\Delta_\vQ = 0$. The white areas are outside the range of the colorbar. The red lines indicate $k_\parallel^2 + k_\perp^2 \pm 2k_\parallel \kf = 0$.}
\end{figure}

\begin{figure}
\includegraphics[width=\columnwidth]{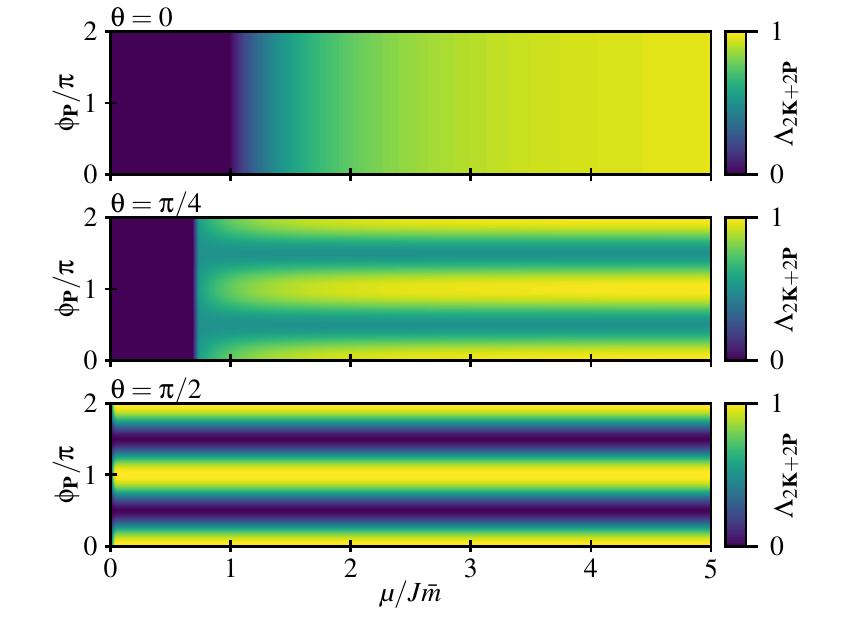}
\caption{\label{fig:LambdaP} Plot of $\Lambda_{2\vK + 2\vP}(0,0)$ for different tilt angles $\theta$ as a function of $\phi_\vP$ and chemical potential $\mu$. The pairing is zero for $\mu < M = J\barm \cos\theta$, since we have no Fermi surface in this case.}
\end{figure}

\begin{figure*}
    \includegraphics[width=1\textwidth]{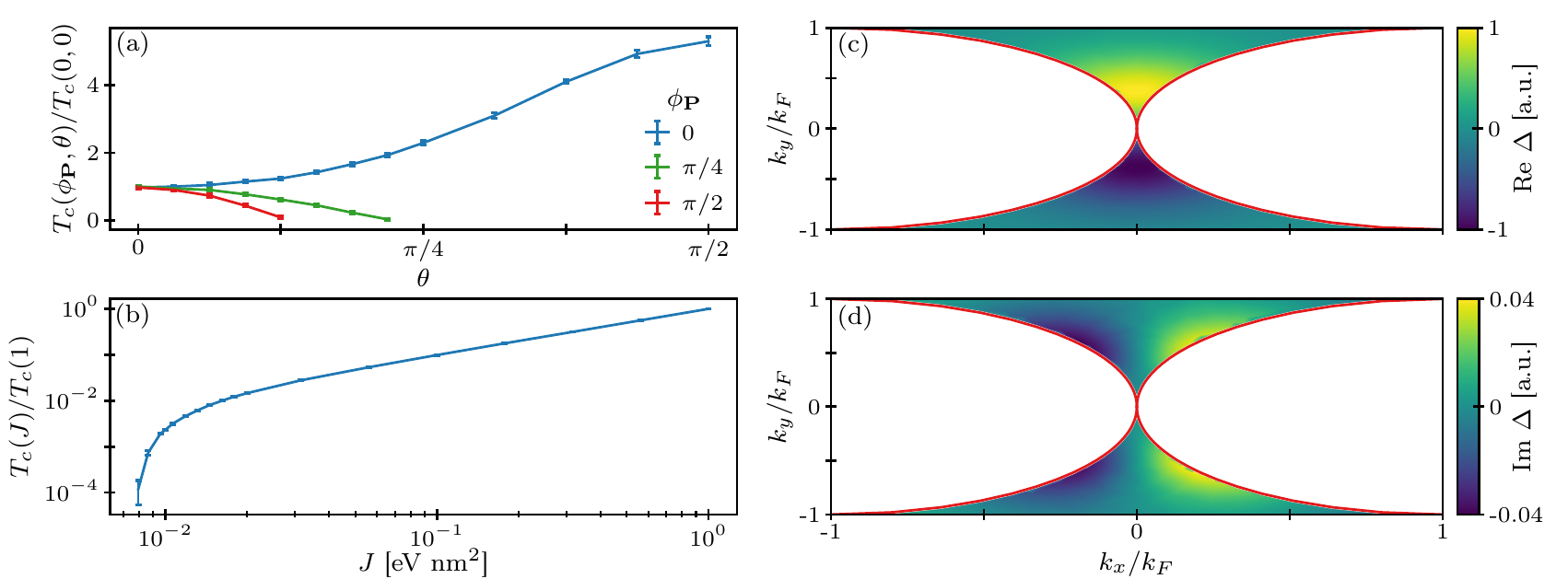}
    \caption{\label{fig:amp_Tc_eigenvecs} (a) Plot of $T_c$ normalized to that at $\phi_\vP = 0$ and $\theta = 0$ for $J = \SI{0.01}{eV nm^2}$. We see that as the tilt angle $\theta$ increases, the critical temperature is no longer the same for different $\vP$: it decreases for $\theta$ increasing towards $\pi/2$ for $\phi_\vP = \pi/4$ and $\pi/2$, as we would expect from \cref{fig:LambdaP}. For $\phi_\vP = 0$ it actually increases for increasing $\theta$. {We use a momentum cut-off of $3\kf$ for $k_x$ and $k_y$.} (b) Plot of $T_c$ as a function of $J$ normalized to the value of $T_c$ at $J=\SI{1}{eV nm^2}$, showing a very sharp decrease in the critical temperature for $J < \SI{0.01}{eV nm^2}$. The error bar shows the sample standard deviation for five calculations of $T_c$. (c) and (d) show the real and imaginary part of the eigenvector $\Delta$ at $T_c$ for $\phi_\vP = 0$, $\theta=0$, and clearly shows that the eigenvector is odd in $\vk$.}
\end{figure*}

As has been shown in previous work \cite{Kargarian2016,Hugdal2018b,Erlandsen2020}, it is possible to get an superconducting instability where the Cooper pairs reside on the same side of the Fermi surface, and the Cooper pairs thus have a finite center of mass momentum of $2\kf$. In the present case, this means setting $\vQ = 2\vK + 2\vP$, where $|\vP| = \kf$. In the limit $T\to 0$, $\chi(\vk)$ quickly drops off to zero when the $\eps_{2\vK+2\vP}^o(\vk)$ term in the $\tanh$ terms dominates over the $\eps_{2\vK+2\vP}^e(\vk)$ term in the $\Delta = 0$ limit, i.e. approximately when
\begin{align}
    k_\parallel^2 + k_\perp^2 \pm 2k_\parallel \kf > 0
\end{align}
where $k_\perp$ ($k_\parallel$) is perpendicular to (parallel with) $\vP$ \cite{Lee2007}, see \cref{fig:chi}.
Even inside this region, we see that $\chi(\vk)$ is largest for small $|\vk|$. In the limit $|\vk|,|\vk'|\ll |\vP|$ the form factor to lowest order is
\begin{align}
    \Lambda_{2\vK + 2\vP}(0,0) = \frac{\vf^2\kf^2(1-\sin^2\phi_\vP\sin^2\theta) + M^2\sin^2\theta}{M^2+\vf^2\kf^2},\label{eq:Lambda_amp}
\end{align}
a plot of which is shown in \cref{fig:LambdaP}. The figure shows that as $\theta$ increases towards $\pi/2$, the isotropy in the $x y$ plane is broken, and pairing of particles with $\vP$ pointing along the $x$ axis become increasingly more favored. Importantly, the sign is opposite compared to the BCS case studied above. Solving the linearized gap equation numerically in the Amperean pairing case as a function of tilt angle $\theta$ for different orientations of $\vP = \kf (\cos\phi_\vP, \sin\phi_\vP)$, we get the results shown in \cref{fig:amp_Tc_eigenvecs}a). As expected from \cref{fig:LambdaP} the critical temperature decreases when $\theta$ increases towards $\pi/2$ when $\phi_\vP = \pi/4$ and $\pi/2$ compared to $\phi_\vP = 0$. For $\phi_\vP = 0$, the critical temperature increases for increasing $\theta$, meaning that Amperean superconductivity might be easier to detect in a system where the FMI magnetization lies in the interface plane. It must be noted that the change in $T_c$ due to changes in $J$ (see \cref{fig:amp_Tc_eigenvecs}b)) is quite large for $J\sim \SI{0.01}{eV nm^2}$, and might explain the rather large relative increase in $T_c$ for $\phi_\vP = 0$ when tuning the magnetization into the plane. Figs.~\ref{fig:amp_Tc_eigenvecs}c) and d) show the real and imaginary part of the eigenvector, showing that the eigenvector is odd in $\vk$. The eigenvector is similar to that obtained in Ref.~\cite{Erlandsen2020} for a topological insulator coupled to an antiferromagnetic insulator.

For systems with a finite in-plane component of the magnetization, the system no longer has many degenerate solutions for all the possible choices of the vector $\vP$. The highest $T_c$ will be for $\vP=(\pm \kf,0)$, and hence we expect the system to condense to either or both of these $\vP$ vectors. Though condensing with $\vP = (\pm \kf,0)$ is equally probable, there is still an overall shift $2\vK$ in the center of mass momentum, meaning we always have a net shift in the Cooper pair center of mass momentum.

\section{Frequency dependent treatment}\label{sec:fullfreq}
We next solve the gap equations including the frequency dependence of the gap function and magnon propagator. In this way we allow for both even-frequency/odd-momentum solutions, and odd-frequency/even-momentum solutions. The latter has, to our knowledge, not been considered in the context of Amperean pairing in other works. Writing out the interaction potential in \cref{eq:basic_gap_eq}, we get
\begin{align}
    \Delta_\vQ^\dagger(i\on,\vk) = {}& \frac{J^2\barm}{\beta V}\sum_{\on',\vk'}\sum_{\gamma} \nonumber\\*
    &\frac{\gamma\omega_{\vk'-\gamma\vk}\Lambda_{\vQ}(\vk',\gamma\vk)\Delta_\vQ^\dagger(i\on',\vk')}{[i\on'-z_1][i\on'-z_2][i\on'-z_\gamma^+][i\on'-z_\gamma^-]},
\end{align}
where $\gamma = \pm 1$, and the poles are given by
\begin{subequations}
\begin{align}
    z_{1,2} ={}& \eps_\vQ^o(\vk') \pm E_\vQ(k'),\\
    z_{\gamma}^\pm ={}& \gamma i\on \pm  2\omega_{\vk' - \gamma\vk}.
\end{align}
\end{subequations}
To find $T_c$ we linearize the above gap equation, and define the indices $N = 2n+1$ and $M = 2n'+1$, and the temperature parameter $t=\pi k_\rm{B} T$, such that the Matsubara frequencies can be written $\on = N t$ and $\on' = M t$.
For notational simplicity, we also define $\Delta_\vQ^\dagger(N,\vk) = \Delta_\vQ^\dagger(i\on,\vk)$. Inserted into the linearized equation, we get
\begin{widetext}
\begin{align}
    \Delta_\vQ^\dagger(N, \vk) ={}& \frac{J^2\barm}{\pi V} t \sum_{M, \vk'} \frac{\Delta_\vQ^\dagger(M,\vk')}{[Mt+i\eps_+(\vk'+\frac{\vQ}{2})][Mt - i\eps_+(-\vk'+\frac{\vQ}{2})]}\nonumber\\*
    &\times\left[\frac{\omega_{\vk'-\vk}\Lambda_{\vQ}(\vk',\vk)}{(M - N)^2t^2 + (2\omega_{\vk'-\vk})^2} - \frac{\omega_{\vk'+\vk}\Lambda_{\vQ}(\vk',-\vk)}{(M + N)^2t^2 + (2\omega_{\vk'+\vk})^2}\right]. \label{eq:gap_eq_NM}
\end{align}
\end{widetext}
Including a finite number $N_\omega$ of positive Matsubara frequencies, and $N_\vk$ reciprocal lattice points $\vk$,  we can write this as a matrix equation $\Delta = \M(t) \Delta$, where $\M(t)$ is a $(2N_\omega N_\vk)\times(2N_\omega N_\vk)$ matrix. Hence, the critical temperature is given by the value of $t$ such that the highest eigenvalue of $\M$ is $1$. 

\begin{figure*}[h!tbp]
    \includegraphics[width=\textwidth]{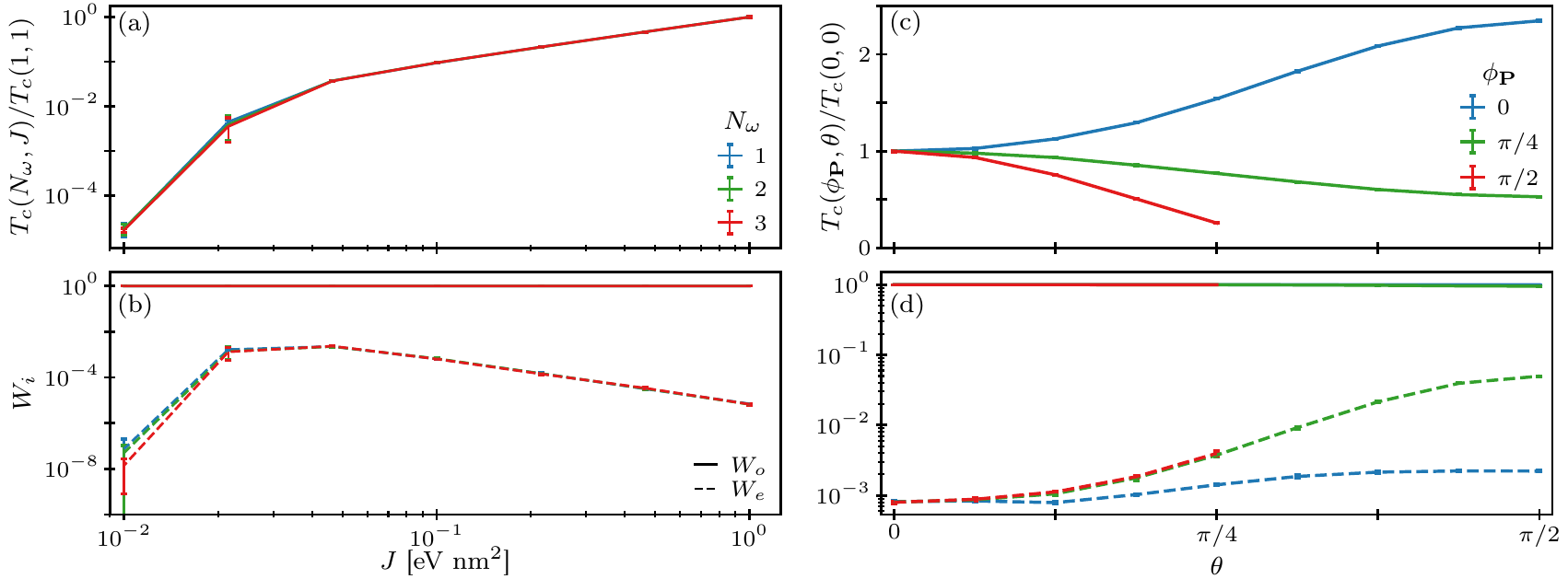}
    \caption{\label{fig:Tc_vs_J} (a) Plot of the critical temperature for different couplings $J$ and $N_\omega$ normalized to that at $J=\SI{e0}{eV nm^2}$ and $N_\omega=1$. (b) Plot of the total weight $W_i$ for odd (solid) and even (dashed) frequency solutions as a function of $J$ for different $N_\omega$. For the entire range of couplings, the eigenvectors are dominantly odd in frequency. For both plots we see that there is no significant difference between the plots for different $N_\omega$. (c) Plot of $T_c$ normalized to that at $\phi_\vP = \theta = 0$ and (d) the total weight $W_i$ as functions of tilt angle $\theta$ for different $\vP$ orientations for $J = \SI{e-1}{eV nm^2}$ and $N_\omega = 1$. The error bars show the sample standard deviation for 5 runs {with momentum cut-off $\kf$}.}
\end{figure*}

Since we did not find any BCS type solutions, except for $\lambda <0$ in the frequency independent treatment above, we will focus only on Amperean pairing.
Solving the eigenvalue problem numerically for the Amperean case with $\vQ = 2\vK + 2\vP$, $\vP = (\kf, 0)$ for $\theta=0$, we find the dependence on coupling $J$ as shown in \cref{fig:Tc_vs_J}a) for $N_\omega = 1$, $2$ and $3$. The critical temperature does not change significantly by increasing the number of Matsubara frequencies included in the calculation. The reason for this is that this is not a strong coupling calculation, and thus the renormalization of the fermion propagator is not included. Hence the largest eigenvalues of $\M$ are given by $M=N=\pm 1$, and necessarily do not change when including more frequencies. 

We also calculate the eigenvalues $\Delta(N,\vk)$ at $T_c$ when solving the matrix equation. Under particle exchange, we must have $\Delta(N,\vk) = -\Delta(-N, -\vk)$ \cite{Linder2019} which means the eigenvectors can be written in the form
\begin{align}
    \Delta(N,\vk) = \Delta_e(N,\vk) + \Delta_o(N, \vk),
\end{align}
where $\Delta_{e/o}$ is even/odd in the frequency index $N$. Hence, we have
\begin{subequations}
\begin{align}
    \Delta_e(N,\vk) ={}& \frac{\Delta(N,\vk) + \Delta(-N,\vk)}{2},\\
    \Delta_o(N,\vk) ={}& \frac{\Delta(N,\vk) - \Delta(-N,\vk)}{2},
\end{align}
\end{subequations}
where $\Delta_{e/o}$ necessarily is odd/even under $\vk\to-\vk$. Numerically we normalize the eigenvectors such that 
\begin{align}
    1 ={}& \frac{1}{V}\sum_{n = -N_\omega}^{N_\omega}\sum_\vk |\Delta(2n+1,\vk)|^2 \nonumber\\*
    ={}& \frac{1}{V}\sum_{n = -N_\omega}^{N_\omega}\sum_\vk \left[|\Delta_e(2n+1,\vk)|^2+|\Delta_o(2n+1,\vk)|^2\right]
\end{align}

For an index $N$, we define the weighting function for odd or even-frequency pairing
\begin{align}
    w_i(N) = {}&\frac{1}{V} \sum_{\vk}\Delta^\dagger_i(N,\vk)\Delta(N,\vk)\nonumber\\*
           = {}&\frac{1}{V} \sum_{\vk}\Delta^\dagger_i(N,\vk)[\Delta_e(N,\vk) + \Delta_o(N,\vk)]\nonumber\\*
           = {}& \frac{1}{V} \sum_{\vk}|\Delta_i(N,\vk)|^2 = w_i(-N)
\end{align}
where $i=e/o$, and the total weight for each symmetry is defined as
\begin{align}
    W_i = \sum_{n=-N_\omega}^{N_\omega} w_i(2n+1).
\end{align}
Hence, we must have
\begin{align}
    1 = \sum_{n=-N_\omega}^{N_\omega} [w_e(2n+1)+w_o(2n+1)] = W_e + W_o.
\end{align}
A plot of $W_i$ is shown in \cref{fig:Tc_vs_J}b), and shows that the odd-frequency part of the eigenvectors dominates the even-frequency part. Hence, this points to the possibility of magnon mediated odd-frequency Amperean pairing, {which is consistent with the fact that pairing at finite momentum has been shown to stabilize odd-frequency superconductivity \cite{Coleman1993,Coleman1993a,Coleman1994}. Moreover, the reason odd-frequency solutions are favored might be understood from the fact that $s$-wave solutions allow for a finite gap close to $\vk=0$, corresponding to a maximum of the first term in \cref{eq:gap_eq_NM}. The even-frequency $p$-wave solution, however, has to be zero at $\vk=0$, and thus gets a much smaller contribution from these areas of $\vk$-space.} 

\begin{figure}[h!tbp]
    \centering
    \includegraphics{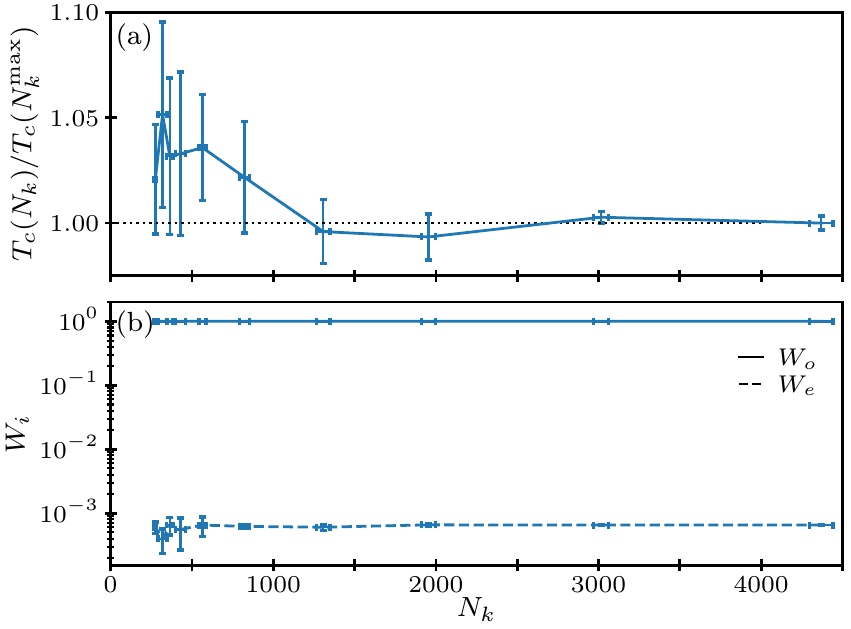}
    \caption{\label{fig:amp_Tc_vs_N} (a) Plot of the critical temperature and (b) weights $W_i$ as functions of the total number of grid points $N_k$ used in the numerical calculation, for $\phi_\vP = \theta =0$, $J = \SI{0.1}{eV nm^2}$, and momentum cut-off $\kf$. The critical temperature is normalized to the value at the highest number of grid points. The figures show an average over 5 runs, with the error bars showing the sample standard deviation. The deviation in the number $N_k$ is due to the way the number of grid points is set by the \protect\UseVerb{term} Python library \cite{Nijholt2019}.}
\end{figure}

Again we see negligible change when increasing $N_\omega$. {\cref{fig:amp_Tc_vs_N} shows the effects of increasing the number of momentum space grid points on the critical temperature and weights $W_i$, showing that $T_c$ converges quickly for the given parameter values. For lower couplings $J$ (not shown), the convergence is slower due to the increasing sharpness of the potential when the temperature decreases. However, the qualitative picture still remains the same independent of the number of grid points, namely that the odd-frequency solution dominates.}

\SaveVerb{term}|adaptive|

Figs.~\ref{fig:Tc_vs_J}c) and d) show the critical temperature and weight functions $W_i$ as functions of tilt angle $\theta$ for different orientations of $\vP$ for $J = \SI{e-1}{eV nm^2}$. The overall $\theta$-dependence is similar to that in \cref{fig:amp_Tc_eigenvecs}a), which is expected since the $\theta$-dependence of $T_c$ is determined by the scattering form factor. Compared to \cref{fig:amp_Tc_eigenvecs}a) the changes in $T_c$ are somewhat less pronounced due to the fact that $T_c$ changes less rapidly as a function of pairing strength in this case, as seen when comparing \cref{fig:Tc_vs_J}a) and \cref{fig:amp_Tc_eigenvecs}b).

\section{Summary}\label{sec:conclusion}
We have derived and solved the gap equation for magnon mediated superconductivity in a TI/FMI bilayer for a general magnetization direction. Neglecting the frequency dependence of the magnon propagator, we found that only Amperean type pairing was possible for easy-axis anisotropy-coupling $\lambda >0$. Tilting the magnetization towards the interface plane lead to an overall shift in the Cooper pair center of mass momenta, and an increase in $T_c$ for Cooper pairs with $\vP$ parallel to the magnetization vector.

Including the frequency dependence of the magnon propagator we found that odd-frequency, even-momentum solutions to the gap equations dominated, thus leading to odd-frequency Amperean pairing. 
{If odd-frequency pairing is found in such a system, it is an example of a naturally occurring odd-frequency superconductor, in contrast to odd-frequency paring due to superconductors coupled to magnetic or spin-orbit coupled materials \cite{Eschrig2011,Linder2015,Eschrig2015a}}
This possibility should be further investigated by performing a strong coupling Eliashberg calculation, where also the frequency dependent renormalization of the fermion propagator is taken into account. {In addition, there are many other properties that should be calculated, such as the Meissner response \cite{Eschrig2015} and the transport properties of the system, which might yield interesting results.}

\begin{acknowledgments}
We acknowledge funding from 
the Research Council of Norway Project No. 250985 "Fundamentals of Low-dissipative Topological Matter", and the Research Council of Norway through its Centres of Excellence funding scheme, Project No. 262633, "QuSpin". H.G.H. thanks E. Erlandsen, E. Thingstad, and J. Linder for useful discussions. Most of the numerics utilized the Python library \protect\UseVerb{term} \cite{Nijholt2019} to generate the $\vk$-space points.
\end{acknowledgments}

\appendix

\section{Details of the calculation of the magnon-mediated interaction}\label{app:interaction}
We rewrite the effective action in \cref{eq:deltaS_TI} in terms of the Dirac fermions defined by \cref{eq:dirac_fermions}. We first get
\begin{align}
    &j(q) = \nonumber\\*
    &\frac{J}{\beta V} \sum_k \begin{pmatrix}
        \Psi_\pm^\dagger(k+q) P_{\vk+\vq}(\cos\theta\s_x - \sin\theta\s_z)P_\vk^\dagger\Psi_\pm(k)\\
        \Psi_\pm^\dagger(k+q) P_{\vk+\vq}\s_y P_\vk^\dagger\Psi_\pm(k)
    \end{pmatrix},
\end{align}
and performing the matrix calculations results in
\begin{align*}
    &P_{\vk+\vq}\s_x P_\vk =\nonumber\\* 
    &\quad\frac{1}{\sqrt{n_\vk n_{\vk+\vq}}}\begin{pmatrix}
        s_\vk r_{\vk+\vq} + s_{\vk+\vq}^* r_\vk &  s_\vk^* s_{\vk+\vq}^* -r_\vk r_{\vk+\vq}\\
        s_\vk s_{\vk+\vq} -r_\vk r_{\vk+\vq} &  - s_\vk^* r_{\vk+\vq} - s_{\vk+\vq} r_\vk
    \end{pmatrix},\\
    &P_{\vk+\vq}\s_y P_\vk =\nonumber\\* 
    &\quad\frac{i}{\sqrt{n_\vk n_{\vk+\vq}}}\begin{pmatrix}
        s_\vk r_{\vk+\vq} - s_{\vk+\vq}^* r_\vk &  -s_\vk^* s_{\vk+\vq}^* -r_\vk r_{\vk+\vq}\\
        s_\vk s_{\vk+\vq} + r_\vk r_{\vk+\vq} &  - s_\vk^* r_{\vk+\vq} + s_{\vk+\vq} r_\vk
    \end{pmatrix},\\
    &P_{\vk+\vq}\s_z P_\vk =\nonumber\\* 
    &\quad\frac{1}{\sqrt{n_\vk n_{\vk+\vq}}}\begin{pmatrix}
        s_\vk s_{\vk+\vq}^* - r_\vk r_{\vk+\vq}  &  - s_\vk^* r_{\vk+\vq} -s_{\vk+\vq}^* r_\vk \\
        - s_\vk r_{\vk+\vq} -s_{\vk+\vq} r_\vk &  -s_{\vk+\vq}s_\vk^* + r_\vk r_{\vk+\vq}
    \end{pmatrix}.
\end{align*}
We will now assume that the chemical potential $\mu> M \ge 0$, meaning that the Fermi level will lie in the $+$--fermion band, and hence only the positive helicity fermions will be free to interact. We therefore keep only the upper diagonal term in the above matrices, resulting in
\begin{widetext}
\begin{align}
    j(q) = \frac{J\barm}{\beta V}\sum_k \frac{\psi^\dagger_+(k+q)\psi_+(k)}{\sqrt{n_\vk n_{\vk+\vq}}} \begin{pmatrix}
        (s_\vk r_{\vk+\vq} + s_{\vk+\vq}^*r_\vk)\cos\theta - (s_\vk s_{\vk+\vq}^* - r_\vk r_{\vk+\vq})\sin\theta\\
        i(s_\vk r_{\vk+\vq} - s_{\vk+\vq}^*r_\vk)
    \end{pmatrix}.
\end{align}
We therefore get, dropping the $+$ subscript on the fields,
\begin{align}
    \delta S_{TI} ={}& - \frac{J^2\barm}{4(\beta V)^3}\sum_{q,k,k'}D(q)\frac{\psi^\dagger(k'+q)\psi^\dagger(k-q) \psi(k)\psi(k')}{\sqrt{n_\vk n_{\vk-\vq} n_{\vk'} n_{\vk'+\vq}}}\nonumber\\*
    &\times\Big[(s_{\vk'}s_{\vk-\vq}^*r_{\vk'+\vq}r_\vk + s_\vk s_{\vk'+\vq}^*r_{\vk'}r_{\vk-\vq})(\cos^2\theta  + 1) + (s_{\vk'}s_\vk s_{\vk'+\vq}^* s_{\vk-\vq}^*  - s_{\vk'}s_{\vk'+\vq}^*r_{\vk}r_{\vk-\vq} \nonumber\\*
    &- s_\vk s_{\vk-\vq}^* r_{\vk'} r_{\vk'+\vq} + r_\vk r_{\vk-\vq} r_{\vk'}r_{\vk'+\vq}-s_{\vk'}s_\vk r_{\vk'+\vq}r_{\vk-\vq} - s_{\vk'+\vq}^* s_{\vk-\vq}^* r_{\vk'}r_\vk)\sin^2\theta\nonumber\\*
    &-(s_{\vk'}s_\vk s_{\vk-\vq}^* r_{\vk'+\vq} + s_\vk s_{\vk'} s_{\vk'+\vq}^* r_{\vk-\vq} + s_{\vk} s_{\vk-\vq}^* s_{\vk'-\vq}^* r_{\vk'}  + s_{\vk'}s_{\vk'+\vq}^* s_{\vk-\vq}^* r_\vk\nonumber\\*
    &- s_{\vk'}r_{\vk'+\vq} r_{\vk} r_{\vk-\vq}  - s_\vk r_{\vk-\vq} r_{\vk'} r_{\vk'+\vq} - s_{\vk'+\vq}^*r_{\vk'}r_{\vk}r_{\vk-\vq} - s_{\vk-\vq}^*r_{\vk}r_{\vk'} r_{\vk'+\vq})\cos\theta \sin\theta \Big]\\
    \equiv& - \frac{J^2\barm}{4(\beta V)^3} \sum_{q,k,k'} D(q)\Lambda_{\vk\vk'}(q) \psi^\dagger(k'+q) \psi^\dagger(k-q) \psi(k)\psi(k'),
\end{align}
where $D(q)$ and $\Lambda_{\vk\vk'}(q)$ are defined in the main text.

\section{Hubbard-Stratonovich decoupling}\label{app:HSdecoupling}
We perform a Hubbard-Stratonovich decoupling \cite{Altland2010} by using the identity
\begin{align}
    1 = \int \D\varphi_q^\dagger \D \varphi_q \exp\left[\beta V\sum_{q,k'k}\varphi_q^\dagger(k')[V_{k'k}(q)]^{-1}\varphi_q(k)\right].
\end{align}
Rescaling the bosonic fields $\varphi_q$,
\begin{align}
    \varphi_q^\dagger(k') \rightarrow{}& \varphi_q^\dagger(k') + \frac{1}{(\beta V)^2}\sum_p \psi^\dagger\left(p+\frac{q}{2}\right) \psi^\dagger\left(-p+\frac{q}{2}\right)V_{pk'}(q),\\*
    \varphi_q(k) \rightarrow{}& \varphi_q(k) + \frac{1}{(\beta V)^2}\sum_p V_{kp}(q)\psi\left(-p+\frac{q}{2}\right) \psi\left(p+\frac{q}{2}\right),
\end{align}
we get
\begin{align}
    \beta V\sum_{q,k',k}\varphi_q^\dagger(k')[V_{k'k}(q)]^{-1}\varphi_q(k) \rightarrow{}& \beta V\sum_{q,k',k}\varphi_q^\dagger(k')[V_{k'k}(q)]^{-1}\varphi_q(k) \nonumber\\*
    &+ \frac{1}{\beta V}\sum_{q,k} \left[\varphi_q^\dagger(k)\psi\left(-k+\frac{q}{2}\right) \psi\left(k+\frac{q}{2}\right) + \varphi_q(k)\psi^\dagger\left(k+\frac{q}{2}\right) \psi^\dagger\left(-k+\frac{q}{2}\right)\right]\nonumber\\*
    &+ \frac{1}{(\beta V)^3}\sum_{q,k',k} \psi^\dagger\left(k'+\frac{q}{2}\right) \psi^\dagger\left(-k'+\frac{q}{2}\right)V_{k'k}(q)\psi\left(-k+\frac{q}{2}\right) \psi\left(k+\frac{q}{2}\right).
\end{align}
Hence we arrive at the functional integral given in \cref{eq:Z}.
\end{widetext}

\section{Material parameters}
\label{app:parameters}

Unless otherwise stated, we have used the parameter values presented in \cref{tab:parameters}.
\begin{table}[h!tbp]
\caption{\label{tab:parameters} Material parameters used unless otherwise stated.}
\begin{ruledtabular}
\begin{tabular}{ll}
$\hbar\vf$ & \SI{0.4}{eV\cdot nm}\cite{Zhang2009}\\
$\mu$ & \SI{0.2}{eV} \cite{Analytis2010}\\
$a$ & \SI{0.4}{nm} \cite{Zhang2009}\\
$J\barm$ & \SI{10}{meV} \cite{Lee2014,Yang2019}\\
$\barm \kappa$ & \SI{0.03}{meV\cdot nm^2} \cite{Bohn1980}\\
$\lambda/\kappa$ & \SI{0.01}{nm^{-2}}
\end{tabular}
\end{ruledtabular}
\end{table}

\bibliographystyle{apsrev4-2}
\bibliography{references}
\end{document}